\documentclass{iau}
\usepackage{graphicx}

\title[The structure and evolution of protoplanetary disks] %% give here short title %%
{The Structure and Evolution of Protoplanetary Disks: \\  an infrared and submillimeter view}

\author[Lucas A. Cieza]   %% give here short author list %%
{Lucas A. Cieza$^1$}

\affiliation{$^1$N\'ucleo de Astronom\'ia,  Universidad Diego Portales, Chile \\email: {\tt lucas.cieza@mail.udp.cl}\\
}

\pubyear{2015}
\volume{314}  %% insert here IAU Symposium No.
\pagerange{119--126}
\jname{Young Stars \& Planets Near the Sun}
\editors{J. H. Kastner, B. Stelzer, \& S. A. Metchev, eds.}
\begin{document}

\maketitle

\begin{abstract}
Circumstellar disks are the sites of planet formation, and the very high incidence of extrasolar planets
implies that most of them actually form planetary systems. Studying the structure and evolution of 
protoplanetary disks can thus place important constraints on the conditions, timescales,
and mechanisms associated with the planet formation process.  
In this review, we discuss observational results from infrared and submillimeter wavelength studies. 
We review disk lifetimes, transition objects,  disk demographics,  and highlight a few remarkable results
 from ALMA Early Science observations. We finish with a brief discussion of ALMA's potential to transform 
the field in near future.


\end{abstract}

\firstsection % if your document starts with a section,

\section{Introduction}

Protoplanetary disks  are complex systems that evolve through various physical mechanisms, including accretion onto the star (Hartmann et al. 1998), grain growth and dust settling
(Dominik, C., $\&$ Dullemond, 2008), dynamical interactions (Artymowicz $\&$ Lubow, 1994),  photoevaporation (Alexander et al. 2006), and planet formation itself (Lissauer 1993; Boss et al. 2000).  
However, the relative importance and timescales of these processes are still not fully  understood.
Describing each of these  processes is beyond the scope of this review.  Instead, we focus on IR and submillimeter\footnote{Submillimeter in this context refers to the
wavelength regime extending from $\sim$0.3 mm to a few mm.}  observational results, while  providing 
a global view of disk evolution and their connection to planet formation theory.  In  \S 2 we discuss results from infrared observations, while in  \S 3 we present submillimeter results previous to 
the commissioning of the Atacama Large Millimeter Array (ALMA). In \S 4 we discuss some of the highlights of ALMA  Early Science observations. 
In \S 5 we speculate on connections between  disk demographics and extrasolar planet statistics.  
In \S 6 we finish with a discussion  on  the prospects for studies with the full ALMA array  and  present our conclusions.  
%3
Complementary discussions on planet formation, chemistry, disk evolution  theory,  and observational studies in other wavelength regimes 
can be found in other contributions to this conference proceedings. 

\section{Infrared constraints}

\subsection{Disk lifetime}

The lifetime of protoplanetary disks is perhaps the strongest astrophysical constraint on planet formation theory, as it provides valuable information on the time available to complete the process. 
For rocky planets, their formation could continue beyond the dispersal of the gas. In the case of giant planets,  the dissipation of the primordial (gas rich) 
disk sets  a hard limit for their formation timescale. 
Since very small amounts of  hot dust (approximately equivalent to an asteroid mass at 1000 K)  are needed to produce an optically thick near-IR (NIR) excess above the stellar photosphere, NIR observations are particularly useful to trace the presence of an inner accretion disk (r $\lesssim$ 0.1 AU).  In fact, there is a very close correspondence between NIR excesses and accretion signatures (e.g. Hartigan et al. 1995),
the only exception being transition disks with completely depleted inner holes. 

From NIR observations of clusters at different ages,  we know that $\sim$80$\%$ of  stars with ages $\sim$1 Myr have an inner accretion disk, and that the inner-disk fraction drops close to 
$\sim$0 $\%$  by 10 Myr (Mamajek, 2009).
From these studies we can derive a mean inner-disk lifetime of 2-3 Myr with a large dispersion: some  pre-main-sequence (PMS) stars accrete for less $<$ 1 Myr, other accrete for up to 10 Myr.   
Mid-infrared (MIR) observations with  \emph{Spitzer} later showed that most targets lacking NIR excesses also lack MIR excesses (Cieza et al. 2007; Wahhaj et al. 2010), again, with the exception of transition disks with clean inner holes.  This indicates that, in general,  once  accretion stops and the inner disk dissipates, the entire disk is dispersed completely rather quickly, in agreement with the predictions of photoevaporation models (e.g., Alexander et al. 2014).  

In any given cluster, disk fractions tend to be lower around higher-mass stars than lower-mass objects (Carpenter et al. 2006), indicating that disk dissipation proceeds faster around  higher mass stars. This could be due to the combination of higher accretion and/or  higher photoevaporation rates. 
Disk lifetime is also a function of multiplicity. In young clusters,  disk frequency is close to 100$\%$ for single stars and much lower for medium-separation (5-50 AU) 
binaries (Kraus et al. 2012; Cieza et al. 2009). This is easy to understand, as mid-separation binaries truncate each other's outer disks, drastically decreasing the amount of material 
available for accretion. Spectroscopic binaries (separation $\lesssim$ 1 AU) and wide binaries (separation $>$ 100 AU) seem to have little effect on disk lifetimes.  

\subsection{Disk structure and evolution}

IR observations are also useful  to derive  disk structures.   Most  protoplanetary disks ($\sim$80$\%$) have ``full disks" extending inward to the dust sublimation radius and are optically thick at all IR wavelengths. As a result, they all have very similar spectral energy distributions (SEDs).  The other 20$\%$, the transition disks, have a wide range of  structures and SEDs.
Protoplanetary disks can be thought of as a series of optically thick annuli at different temperatures. Each annulus dominates the emission at a different wavelength; the closer to the star and the hotter the annulus, the shorter the corresponding wavelength. Therefore, the presence of an inner hole in the disk translates to a reduced level  of NIR excess.  If the hole is completely empty,  no detectable excess will be seen above the stellar photosphere.   If some detectable dust remains inside the cavity, an optically thin excess will still be present, but the NIR SED will stay below the typical value of an optically thick inner-disk.  Similarly, if the disk contains a wide enough gap that is optically thin in the IR, its SED will show a dip at the wavelength corresponding  to the temperature of the ``missing" dust annulus. 
Furthermore, most young protoplanetary disks are flared  and thus intercept and reprocess significantly more starlight than a flat disk. As disks evolve and dust  grows  and settles to the mid-plane,  the outer disk becomes flatter with time, which translates to reduced levels of mid- and far-IR  emission with respect to a fully flared disk.  
Finally, as the primordial disk dissipates, it becomes optically thin at all IR wavelengths, resulting in reduced levels of IR excess at all wavelengths. 

In practice, most young stellar clusters and star-forming regions contain two categories of disks: normal  ``full" disks that are optically thick and transition disks. The latter group  includes objects with inner holes and gaps 
(sometimes called classical transition disks, cold disks, or pre-transition disks) and objects that are physically very  flat or optically thin in the IR (some times called  anemic, weak-excess or homologously depleted disks)\footnote{For a discussion  on disk nomenclature,  see the \emph{Diskionary}  by  Evans et al. (2009).}.
The relative number of these types of objects seems to be a function of age, with more transition disks in older  star-forming regions and clusters (Currie et al. 2009). 

\section{Pre-ALMA submillimeter results} 

\subsection{Disk mass}

While most protoplanetary disks remain optically thick in the IR, they become optically thin at longer wavelengths. Submillimeter wavelength
observations are hence sensitive to fundamental disk properties such as total gas and dust masses and grain size distributions, and 
are highly complementary to IR data.
The main constituent of protoplanetary disks, H$_2$, is a very poor emitter under the relevant temperatures and
densities. Circumstellar dust is much easier to observe from continuum observations. Estimating
the mass of a disk, and hence its ability to form different types of planets, typically entails making very strong
assumptions about the dust opacity and the gas to dust mass ratio (Williams $\&$ Cieza, 2011). 
Submillimeter fluxes are  routinely used to estimate the masses of protoplanetary disks using:

\begin{equation}
M_{dust}  =  \frac{F_\nu d^2}{\kappa_\nu B_\nu (T_{dust})}
\end{equation}

\vspace{0.3cm}
\noindent where $d$ is the distance to the target,  $T$ is the dust temperature
and  $\kappa_{\nu}$ is the dust opacity.  
Following Beckwith et al.  (1990), 
and making  standard  (although uncertain) assumptions about  the disk temperature (T$_{dust}$ = 20 K), the  dust opacity ($\kappa_{\nu}$ = 10[$\nu/$1200 GHz] cm$^2$g$^{-1}$),  
and the gas to dust mass ratio (100), Equation 3.1  becomes:

\begin{equation}
M_{disk} (gas + dust)  =  1.7  \times 10^{-4}  \left( \frac{F_{\nu}(1.3 mm)}{mJy}  \right) \times   \left(  \frac{d}{140 pc}  \right) ^2  M_{\odot}
\end{equation}
\vspace{0.3cm}

\vspace{0.3cm}
The gas to dust mass ratio of 100 that is usually adopted is appropriate for the interstellar medium, but highly questionable for protoplanetary disks that are known to undergo significant
grain growth and photoevaporation.  In fact, understanding the evolution of the gas (and the gas to dust mass ratio) in protoplanetary disks remains one of the main challenges in the 
fields of disk evolution and planet formation. The second most common molecule after H$_2$ is CO, and its isotopologues, $^{13}$CO and C$^{18}$O, provide a viable approach to estimate 
the gas content in disks (Williams $\&$ Best,  2014).

From continuum submillimeter surveys of nearby star-forming regions such as Taurus and Ophiuchus, the following benchmark properties have been derived:   
M$_{disk}$  $\propto$ M$_{\star}$ 
and 
M$_{disk}$ $\sim$0.5$\%$ M$_{\star}$
 (Andrews $\&$ Williams  2005, 2007; Andrews et al. 2013). However, the dispersion is large ($\pm$ 0.7 dex) and 
the submillimeter luminosities of protoplanetary disks in the same regions, and with almost identical IR SEDs,  can differ by up to two orders of magnitude 
(e.g., Cieza et al. 2008; 2010),  implying drastically different disk masses and/or grain-size  distributions.
Understanding disk evolution requires both IR and submillimeter observations.  Unfortunately,  millimeter surveys of disks in molecular clouds and young stellar clusters clearly lag far behind their IR counterparts. While deep and complete IR censuses of disks exist for tens of regions spanning a wide range of ages and IR disk fractions,  Taurus is the only region in which the entire disk population has been observed at submillimeter wavelengths with enough depth to  detect the majority of the IR-detected disks  (Andrews et al. 2013).

\subsection{Resolved observations}

Using interferometers and imaging synthesis techniques, nearby protoplanetary disks can be spatially resolved, providing direct information on the surface density profile of the disk, which is critical for planet formation theory.  Simultaneous modeling of the SED and submillimeter images is a powerful technique to constrain the basic properties of disks.  
Thus far, resolved submillimeter studies have been limited to continuum observations of  bright sources  (F$_{mm}$ $>$50 mJy) and/or gas observations of optically thick lines. Therefore, they are very biased towards massive
disks around relatively massive stars and provide little direct information on the gas content of disks. In particular, most imaging studies have focused on transition objects (see Williams $\&$ Cieza, 2011 for a review). 
These early resolved observations indicated that the surface density profiles in many of the bright protoplanetary disks are consistent with the amount of material needed for the formation of the planets in the Solar System (Andrews et al. 2010). 
With the commissioning of ALMA, we should be able to 1) extend the imaging surveys to less massive,  more typical,  protoplanetary disks and 2) image the disks in optically thin gas tracers 
such as $^{13}$CO and C$^{18}$O in order to trace the surface density profile of the gas more directly.

\section{ALMA Early Science results}

Since the start of Early Science observations in 2011, ALMA is shedding new light on every stage of disk evolution, from the deeply embedded Class I stage\footnote{The Class I stage corresponds to a disk that is still embedded in its natal envelope.} to the dissipation of the primordial disk and the transition to the debris disk phase. In the following, we highlight some of remarkable ALMA Early Science results.

\subsection{Rings in Class I circumstellar disks}

Long-baseline ALMA observations at 0.025$"$ to 0.075$"$ (3.5 -- 10 AU) resolution of the Class I  circumstellar disk around HL Tau revealed a set of  impressive concentric gaps. These gaps  show evidence for grain growth, are slightly eccentric, and many of them appear to be in resonance with each other (Partnership et al. 2015). All these features are highly suggestive of  advanced planet formation through core accretion, although other explanations have merit,  including the growth of pebbles at  the snow lines of different species (Zhang et al. 2015).   If the gaps in the HL~Tau disk are in fact due to orbiting planets massive enough to carve them, it would imply that the core accretion mechanism is significantly more efficient that previously thought  and that the planet formation process can be fairly advanced by the Class I stage  and an age of  $\sim$1 Myr. 
 
\subsection{Structures in transition objects}

High-resolution Early Science observations of transition disks have also produced some surprises such as the discovery of huge asymmetries in the outer disks of  Oph~IRS ~48 (van der Marel et al. 2013),  SR~21 (Perez et al. 2014 ), and HD~142527 (Casassus et al. 2013). These asymmetries have been interpreted as large-scale vortices or dust traps, which could play an important role in the formation of planetesimals and planets at large radii.  HD~142527 also shows a set of spiral arms (Christiaens et al. 2014) of unknown origin and gaseous accretion flows connecting the inner and the outer disk components of the system (Casassus et al. 2013). Such accretion flows have been predicted by hydrodynamical simulations of multiple planets embedded in a primordial disk (Dodson-Robinson 
$\&$ Salyk, 2011). All these results demonstrate that transition disks have complex structures and that  the models derived from SED fitting and low-resolution images can only be considered  zero-order approximations to their true structures. 

\subsection{The primordial to debris disk transition}

As discussed in \S 2.2,  star-forming regions include very diverse types of young stellar objects: ``full" disks, disks with inner-holes and gaps,  optically thin disks, and diskless stars. 
The evolutionary status of the optically thin disks has remained a matter of intense debate over the last few years and these objects have often been considered the final stage of the primordial disk phase:
evolved,  primordial (gas rich) disks where most of the primordial dust has been depleted.  
In an ALMA survey of 24 non-accreting PMS stars (weak-line T Tauri stars; WTTSs) with known IR excesses, Hardy et al. (2015) found that these systems have  dust masses 
$\lesssim$ 0.3 M$_{\oplus}$  and no evidence of gas, and  therefore are more akin to young debris disks than to evolved primordial disks.  This would indicate that they are already in a debris stage, as opposed to \emph{transitioning} into that phase. 
Non-accreting PMS with IR excesses represent  $\sim$20$\%$ of the WTTS population (Cieza et al. 2007), and their incidence has been used to estimate  the dissipation timescale of the primordial disks once accretion stops and Classical T Tauri stars (CTTSs)  evolve into WTTSs  (Wahhaj et al. 2010). If WTTS disks are produced by second-generation dust, it would imply that the final inside out dispersal of the primordial disk  happens even faster than previously thought ($\sim$1$\%$ vs $\sim$10$\%$ of the disk lifetime).  These ALMA results also raise questions on how to initiate the debris disk phenomenon within a few Myr of  the formation  of the star and why this is seen in $\sim$20$\%$ of the PMS stars that have already lost their primordial disks.

\section{Disk evolution and planet formation}

It is usually assumed that ``full disks" are the starting point of disk evolution and that  the transition disks with holes and gaps are more ``evolved"  systems. However, demographic studies (Cieza et al. 2010; Romero et al. 2012) in nearby star-forming regions suggest that there are at least two different evolutionary paths that disks can follow. Some systems develop gaps and inner-holes while their disks are relatively massive and still accreting,  while other objects lose most of their disk mass while keeping  a perfectly ``normal" IR SED for millions of years  until they reach a point in which the entire disk 
dissipates from the inside out in a small fraction of the disk lifetime (most likely through photoevaporation). 
The former group includes most of the famous transition disks found in the literature. In general, these are early-type stars (early-K to A-type) have massive disks with enough material to form the Solar System and large inner-holes, again, consistent with the size of the Solar System.  Despite their high profile in the literature, such systems are \emph{not} the typical young stellar object in a nearby molecular cloud or even the typical transition disk. Instead, they represent a minority of the order of  20$\%$ of the transition disk population, and  5$\%$ of the general disk population (Cieza et al. 2012). It is tempting to speculate that such systems are the minority of disks that form planetary systems with one or more giant planets. 
On the other hand, the ``typical" disk in a young (2-3 Myr) stellar cluster seems to be a low-mass  accreting CTTS with a normal SED, but not enough mass left in the disk to form a giant planet (Cieza et al. 2015).  Such systems can be considered ``evolved" in the sense that they have probably already lost most of their initial disk mass and  have also undergone significant grain growth and dust settling (Lada et al. 2006).  Given the fact that rocky planets are ubiquitous in the Galaxy (Batalha et al. 2013), it is also tempting to speculate that these typical disks are the ones that form the more typical planetary systems, i.e., those hosting rocky planets but no giant planets.
 
\section{Prospects for full-ALMA observations and conclusion}  

With unprecedented sensitivity and resolution,  ALMA is clearly poised to revolutionize the fields of disk evolution and planet formation in the near future. It will have an impact in all stages of disk evolution. 
We can expect to see many detailed studies of disks in early stages, when they still remain embedded in their natal envelope. High resolution images at long wavelengths and the use of optically thin lines allow us to  zoom in and peek inside the envelope emission and reveal the disk structure.  Some of the questions that the full ALMA array will address include: are rings like those of HL Tau common in Class I objects? What is their origin? Are very young protoplanetary disks gravitationally unstable? Can we image collapsing clumps in the process of forming brown dwarfs or giant planets? We can also expect to see many more detailed studies of transition systems which will investigate the origin of their intriguing features (e.g., large cavities, spiral arms, asymmetries, clumps) and their connection to planet formation. However, as discused in \S 5, such systems are not typical and might be related to the formation of some particular types of planetary systems. 
If we want to learn how more common planetary systems form, we will also need to study more common protoplanetary disks. In this regard, ALMA's sensitivity will make it possible to perform complete surveys and resolve entire disk populations in nearby molecular clouds using both continuum and gas tracers. This will help us to investigate the full distribution of disk properties (gas and dust masses, sizes, surface density profiles,  etc.)  as a function of stellar parameters such as mass, age,  and multiplicity. In particular, the observations of optically thin line tracers such as $^{13}$CO and C$^{18}$O will allow us to place much needed constraints on the evolution of gas, which is far less established than that of the dust.  
Thanks to the exponential growth of extrasolar planet studies and the construction of ALMA,  we are approaching an era in which we can start making direct connections between disk properties and the statistics of planetary systems we see in the Galaxy. This will help us to understand the kind of disks that form different types of planetary systems and to develop a coherent picture of the demographics of protoplanetary disks and the planets they form.

\end{document}